\documentclass{iau} 
\usepackage{graphicx}

\newcommand\apj{ApJ}
\newcommand\apjl{ApJL}

\newcommand\mnras{MNRAS}
\newcommand\aap{A\&A}

\title[Magnetar formation] 
{How to form a millisecond magnetar? Magnetic field amplification in protoneutron stars}

\author[Guilet et al.]   
{J\'er{\^o}me Guilet$^{1,2,3}$, Ewald M{\"u}ller$^1$, Hans-Thomas Janka$^1$, Tomasz Rembiasz$^4$, Martin Obergaulinger$^4$, Pablo Cerd\'a-Dur\'an$^4$, Miguel-Angel Aloy$^4$
}

\affiliation{$^1$Max-Planck-Institut f{\"u}r Astrophysik, Karl-Schwarzschild-Str. 1, D-85748 Garching, Germany \\[\affilskip]
$^2$Max-PlanckÐPrinceton Center for Plasma Physics \\[\affilskip]
$^3$Laboratoire AIM, CEA/DRF-CNRS-Universit{\'e} Paris Diderot, IRFU/D{\'e}partement d'Astrophysique, CEA-Saclay, F-91191, France\\[\affilskip]
$^4$Departamento de Astronom'a y Astrof'sica, Universidad de Valencia C/ Dr. Moliner 50, 46100 Burjassot, Spain}

\pubyear{2017}
\volume{331}  
\jname{SN 1987A 30 Years Later}
\editors{M. Renaud, A. Markowith, G. Dubner, A. Ray \& A. Bykov eds.}
\begin{document}

\maketitle

\begin{abstract}
Extremely strong magnetic fields of the order of $10^{15}\,{\rm G}$ are required to explain the properties of magnetars, the most magnetic neutron stars. Such a strong magnetic field is expected to play an important role for the dynamics of core-collapse supernovae, and in the presence of rapid rotation may power superluminous supernovae and hypernovae associated to long gamma-ray bursts. The origin of these strong magnetic fields remains, however, obscure and most likely requires an amplification over many orders of magnitude in the protoneutron star. One of the most promising agents is the magnetorotational instability (MRI), which can in principle amplify exponentially fast a weak initial magnetic field to a dynamically relevant strength. We describe our current understanding of the MRI in protoneutron stars and show recent results on its dependence on physical conditions specific to protoneutron stars such as neutrino radiation, strong buoyancy effects and large magnetic Prandtl number.
\end{abstract}

\firstsection 
\section{Introduction}
The delayed injection of energy due to the spin down of a fast rotating, highly magnetized neutron star is the most popular model to explain a class of superluminous supernovae (e.g. Kasen \& Bildsten 2010; Inserra et al. 2013). The birth of such Ómillisecond magnetarsÓ is furthermore a potential central engine for long gamma-ray bursts and hypernovae (e.g. Duncan \& Thompson 1992; Metzger et al. 2011; Obergaulinger \& Aloy 2017). One of the most fundamental and open question in these models is the origin of the strong magnetic field that is invoked. The most studied mechanism capable of generating such a strong magnetic field is the growth of the magnetorotational instability (MRI) in the protoneutron star (Akiyama et al. 2003; Masada et al. 2007; Obergaulinger et al. 2009; Sawai \& Yamada 2014; Guilet et al. 2015; Guilet \& M\"uller 2015; Rembiasz et al. 2016a,b; M\"osta et al. 2016). In the following we review several aspects of the physics of the MRI in the physical conditions relevant to protoneutron stars, by describing its linear growth (Section~2) and its non-linear dynamics (Section~3).

\section{Linear growth of the MRI}
\begin{figure}[htbp]
\begin{center}
 \includegraphics[width=11cm]{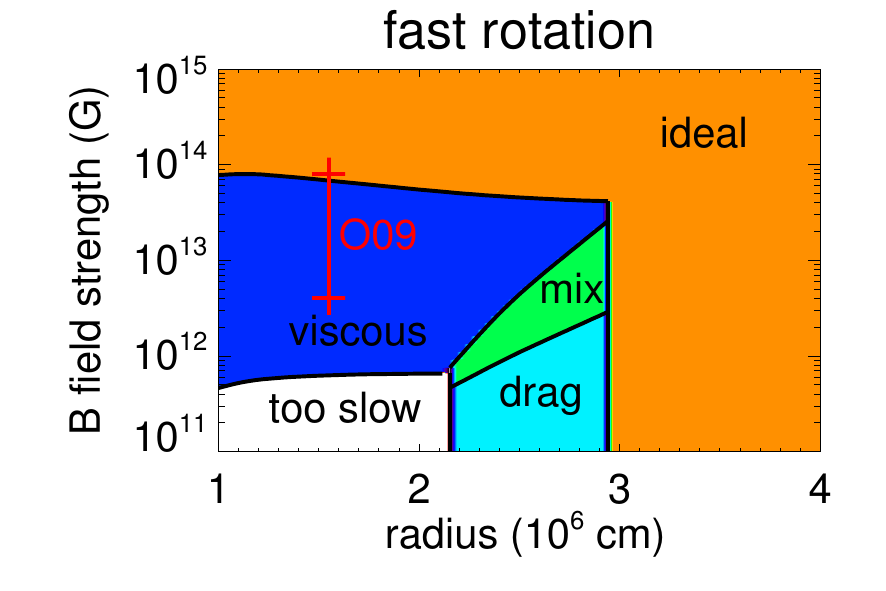} 
 \caption{Different regimes of MRI growth as a function of radius and magnetic field strength in a fast rotating PNS model at $t=170\,{\rm ms}$ post-bounce (from Guilet et al. 2015). See the text for a description of the different regimes. The parameter range used in the simulations by Obergaulinger et al. (2009) is shown in red. The regime labelled as "too slow" corresponds to growth rates lower than $10\,s^{-1}$.}   \label{fig:MRI_regimes}
\end{center}
\end{figure}

\begin{figure}[htbp]
\begin{center}
 \includegraphics[width=6.7cm]{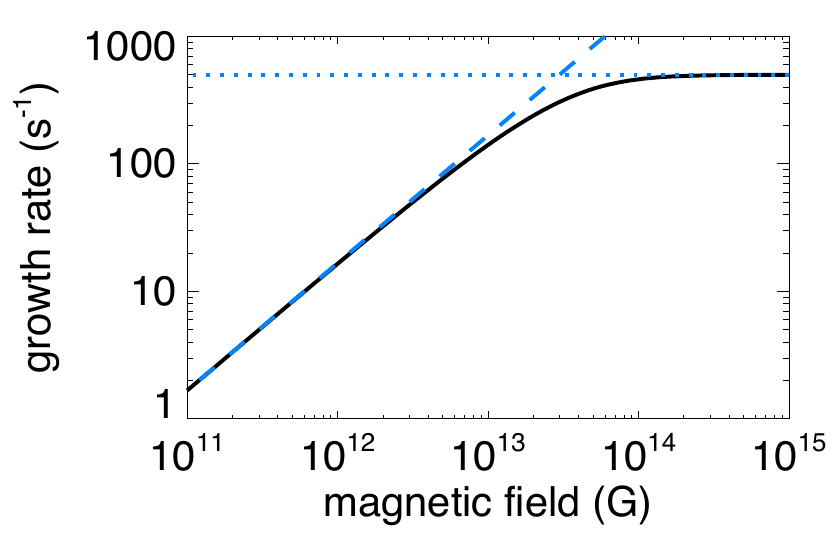} 
  \includegraphics[width=6.7cm]{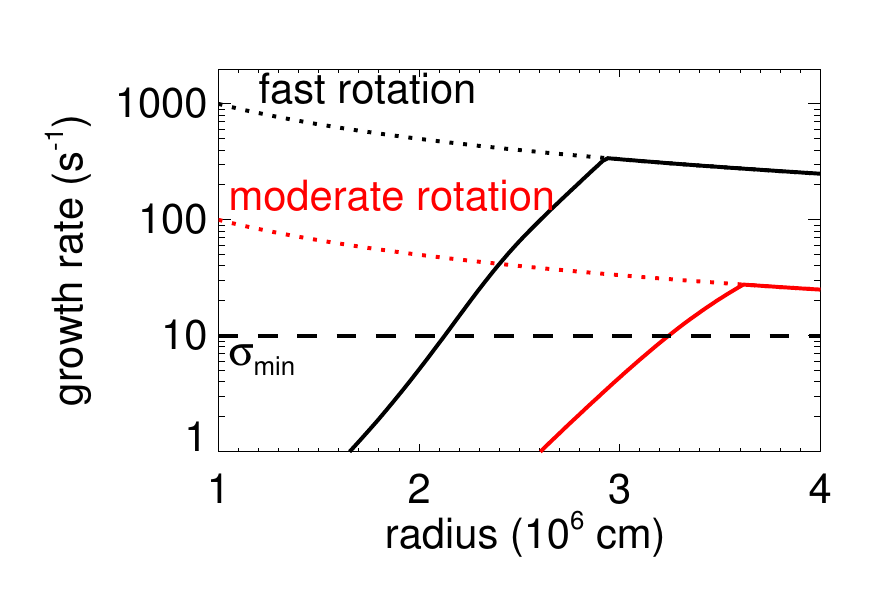} 
 \caption{Maximum growth rate of the MRI in the viscous regime (left panel) and in the drag regime (right panel). \textit{Left panel:} the solid line shows the MRI growth rate taking into account neutrino viscosity, while the dotted line shows the ideal MHD limit. The dashed line shows equation~(15) of Guilet et al. (2015), which is valid in the very viscous regime ($E_\nu\ll 1$).  \textit{Right panel:} Solid lines show the MRI growth rate including the impact of neutrino drag, while dotted lines show the growth rate in the ideal MHD limit. Black and red colors represent, respectively, a profile of fast rotation and a ten times slower rotation (see Guilet et al. 2015 for details). Finally, the dashed line shows the (somewhat arbitrary) minimum growth rate below which the growth is labelled as "too slow" in Fig.~\ref{fig:MRI_regimes}.}
   \label{fig:MRI_sigma}
\end{center}
\end{figure}

Although it is too often neglected in numerical simulations, neutrino radiation can have a dramatic impact on the linear growth of the MRI. This was shown by Guilet et al. (2015) by applying analytical results for the linear growth of the MRI to a numerical model of a protoneutron star (PNS).
In Guilet et al. (2015), we have shown that, depending on the physical conditions, the MRI growth can occur in three main different regimes (Fig.~1):
\begin{itemize}
\item {\underline{\it Viscous regime}} (dark blue color in Fig.~\ref{fig:MRI_regimes}) : on length-scales longer than the neutrino mean free path, neutrino viscosity significantly affects the growth of the MRI if $E_\nu\equiv \frac{v_{A}^2}{\nu\Omega}<1$, where $v_A$ is the Alfv\'en velocity, $\nu$ the viscosity induced by neutrinos and $\Omega$ the rotation angular frequency. The growth of the MRI is then slower and takes place at longer wavelengths compared to the ideal regime. In the viscous regime, the wavelength of the most unstable mode is independent of the magnetic field strength, while the growth rate is proportional to the magnetic field strength (Fig.~\ref{fig:MRI_sigma}). As a result, a minimum magnetic field strength of $\sim 10^{12}\,{\rm G}$ is required for the MRI to grow on sufficiently short time-scales.
\item {\underline{\it Drag regime}} (light blue color in Fig.~\ref{fig:MRI_regimes}) : on length-scales shorter than the neutrino mean free path, neutrino radiation exerts a drag force $-\Gamma \bf{v}$, where $\bf{v}$ is the fluid velocity perturbation and $\Gamma \simeq 6\times 10^3\,(T/10\,{\rm MeV})^6\,{\rm s^{-1}}$ is a damping rate. This drag has a significant impact on the MRI if the damping rate is larger than the rotation angular frequency ($\Gamma > \Omega$). In this regime, the growth rate of the most unstable mode is independent of the magnetic field strength, but is reduced by a factor of $\Gamma/\Omega$ compared to the ideal regime (Fig.~\ref{fig:MRI_sigma}). The wavelength of the most unstable mode is not much affected by the neutrino drag.   
\item {\underline{\it  Ideal regime}} (orange color in Fig.~\ref{fig:MRI_regimes}) : this is the classical MRI regime which occurs when neutrino viscosity or drag are negligible. The growth rate of the MRI is then a fraction of the angular frequency, independent of the magnetic field strength, while the most unstable wavelength is proportional to the magnetic field strength.
\end{itemize}

Fig.~\ref{fig:MRI_regimes} shows where these three regimes apply, as a function of magnetic field strength and radius in the PNS (note that qualitatively similar results were obtained by Guilet et al. (2017) in the context of neutron star mergers). Three regions in the PNS can be distinguished:
\begin{itemize}
\item Deep inside the PNS, the relevant MRI regime is the viscous MRI. The MRI can grow on sufficiently short time-scales if the initial magnetic field is above a critical strength.
\item At intermediate radii, MRI growth can take place both in the viscous regime at wavelengths longer than the neutrino mean free path, and in the drag regime at length-scales shorter than the mean free path. Since the growth rate in the viscous regime is proportional to the magnetic field strength, the growth is faster in the viscous regime above a critical magnetic field strength while it is faster in the drag regime for weaker magnetic fields. In addition, in-between these regimes, the MRI growth occurs in a mixed regime (shown in green in Fig.~\ref{fig:MRI_regimes}) where electron neutrinos are diffusing and thus induce a viscosity, while the other species are free streaming and exert a drag.
\item Near the PNS surface, the viscous regime is irrelevant because the neutrino mean free path is longer than the wavelength of the MRI. Furthermore, in this region the neutrino drag does not affect much the growth of the MRI because the damping rate is smaller than the angular frequency, i.e. $\Gamma<\Omega$. As a consequence the MRI growth takes place in the ideal regime without much impact of neutrino radiation.
\end{itemize}

For the sake of simplicity and in order to show the impact of neutrinos in a clear way, this analysis neglected the effect of buoyancy. As shown analytically by Menou et al. (2004) and Masada et al. (2007) and confirmed by numerical simulations in Guilet \& M\"uller (2015), entropy and composition gradients can act against the MRI, but this is alleviated by the diffusion due to neutrinos.

\section{Non-linear dynamics of the MRI}
In order to know the efficiency of magnetic field amplification, numerical simulations of the non-linear dynamics driven by the MRI are necessary. These simulations have so far mostly been performed in local or semi-global models describing a small portion of the PNS (e.g. Obergaulinger et al. 2009; Masada et al. 2012; Guilet et al. 2015; Rembiasz et al. 2016a,b). The first phase of MRI growth is often dominated by channel flows, which are the fastest growing MRI modes in the presence of a poloidal magnetic field. These modes, which are uniform in the horizontal direction, are approximate non-linear solutions and are therefore potentially able to grow well into the non-linear regime. Their growth is terminated when parasitic instabilities (either the Kelvin-Helmholtz instability or resistive tearing modes) are able to destroy their structures (Goodman \& Xu 1994; Rembiasz et al. 2016a,b). By studying in detail this process with two different numerical codes, Rembiasz et al. (2016b) have been able to show that channel modes can only amplify the magnetic field by a factor $\sim 10$ under the physical conditions prevailing in a PNS. A further process such as a turbulent MRI-driven dynamo is therefore necessary to reach very strong magnetic fields. 

\begin{figure}[htbp]
\begin{center}
 \includegraphics[width=6.7cm]{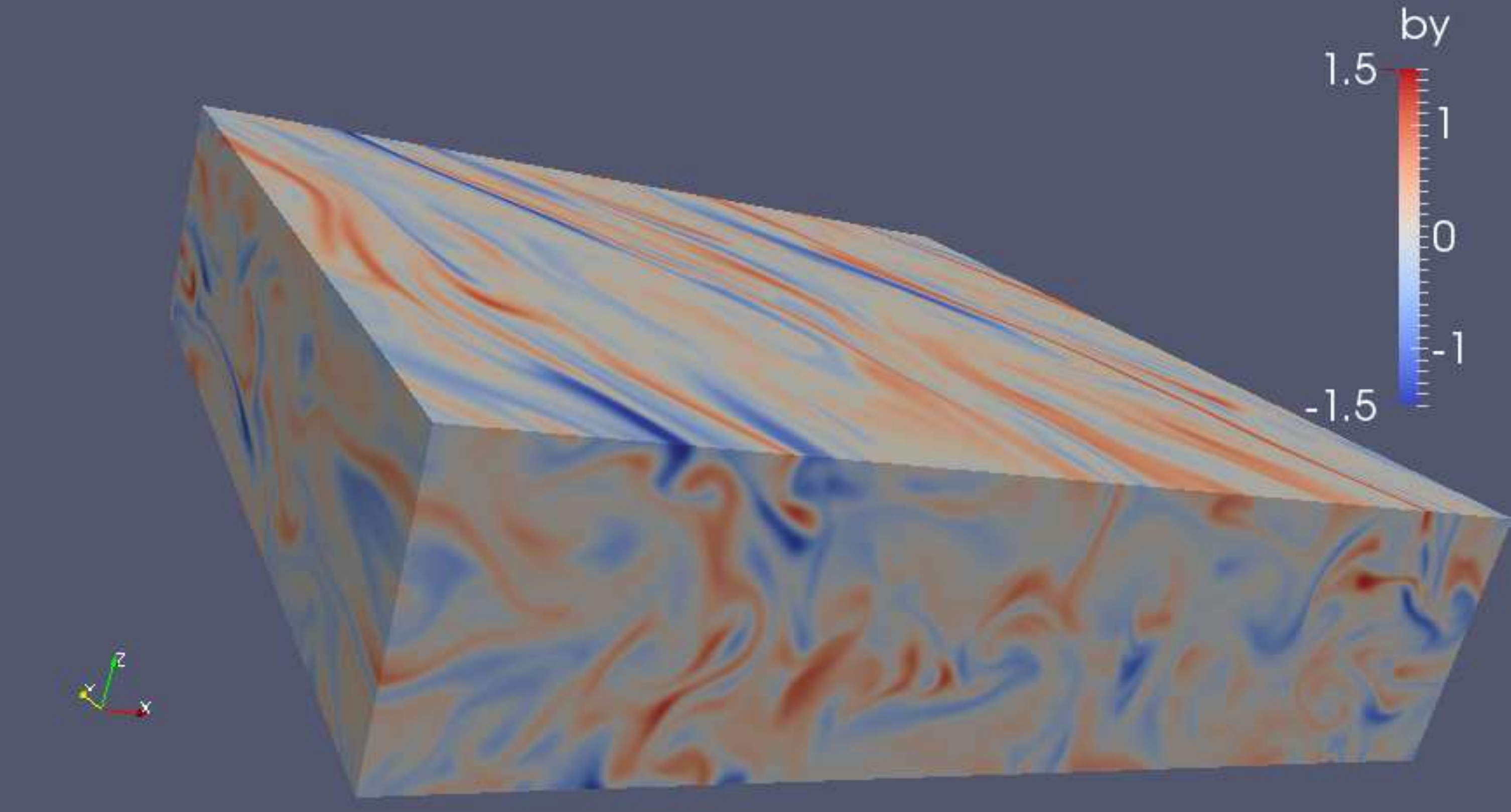} 
 \includegraphics[width=6.7cm]{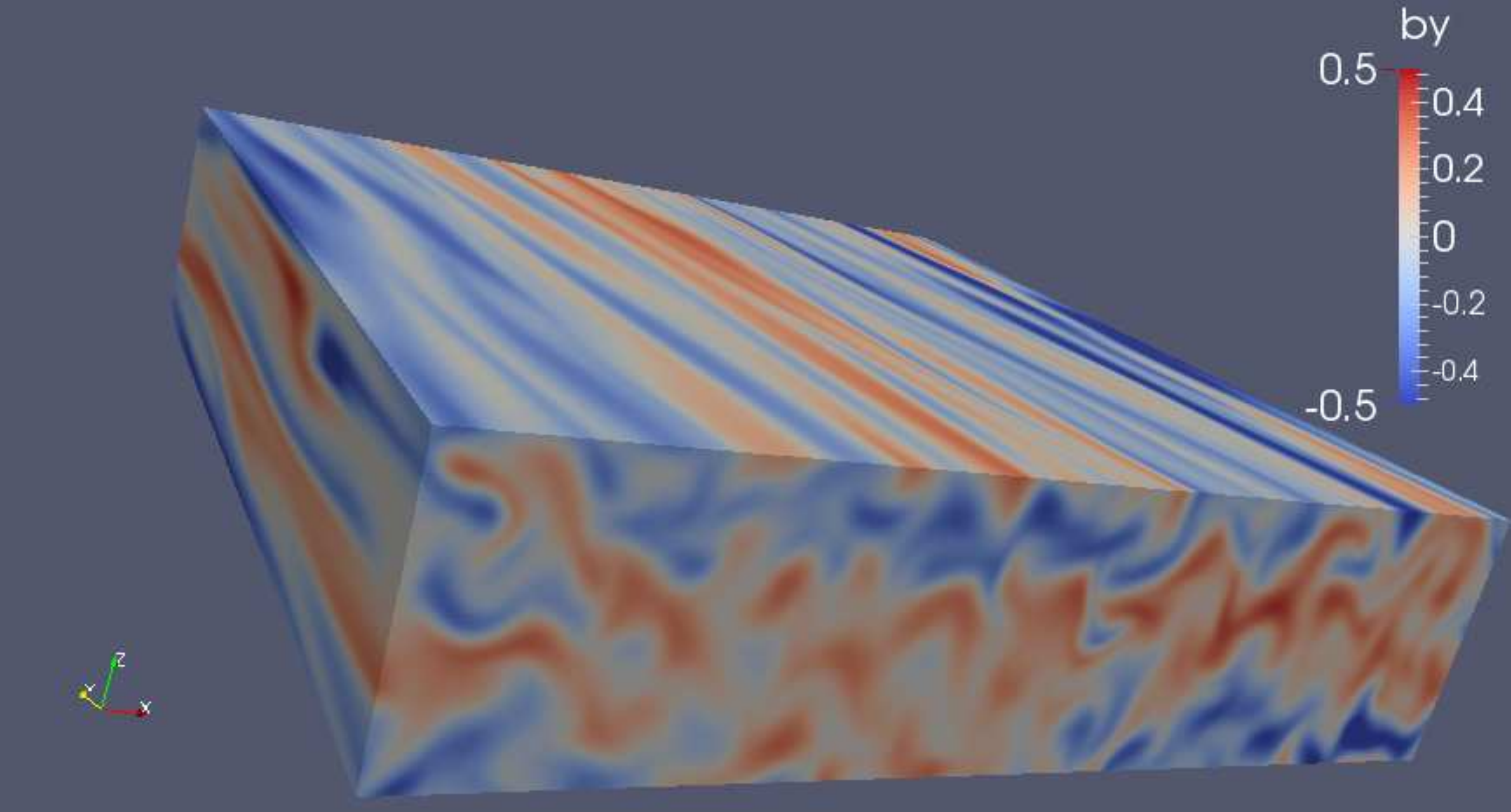} 
  \includegraphics[width=6.7cm]{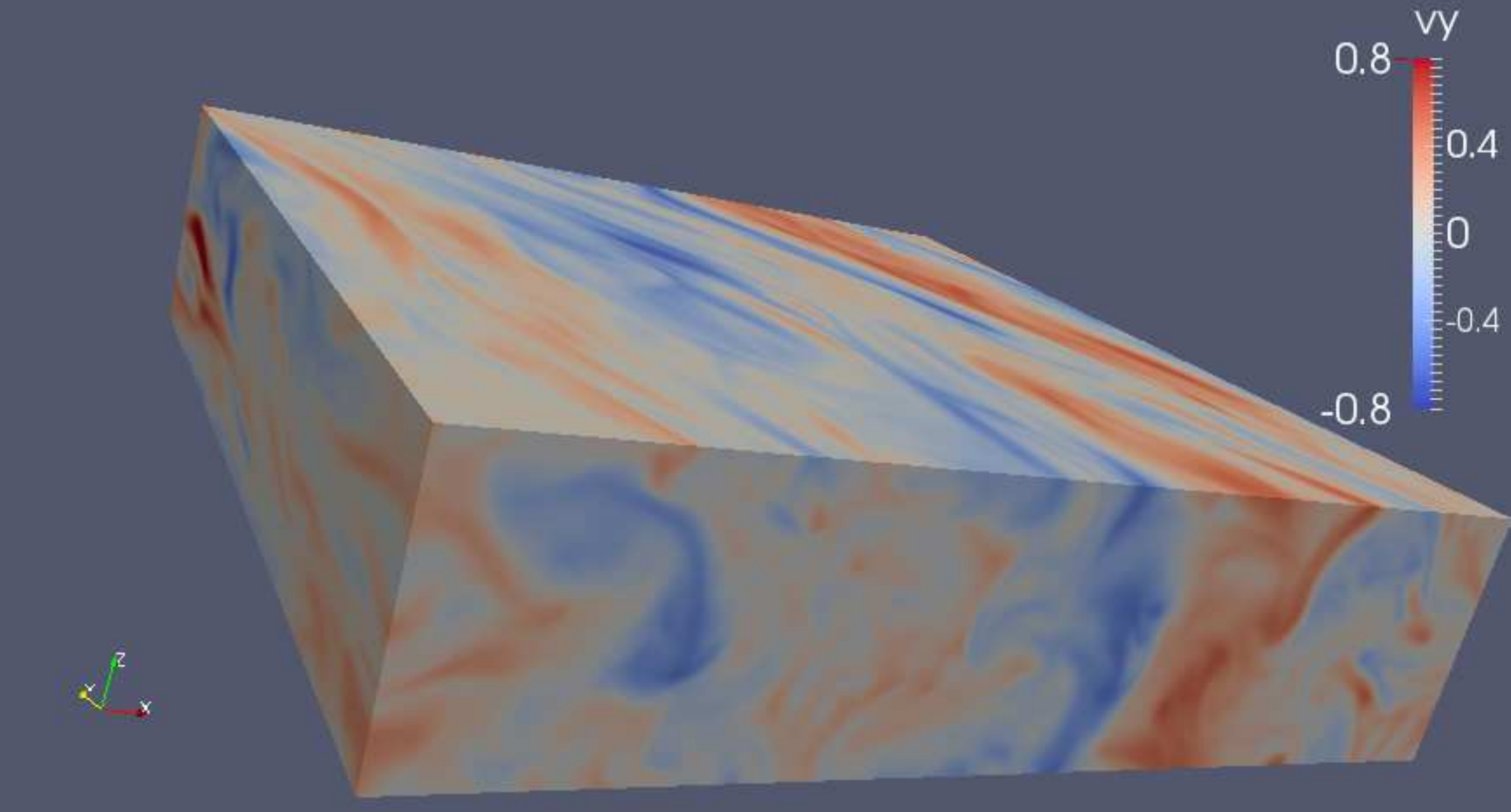} 
 \includegraphics[width=6.7cm]{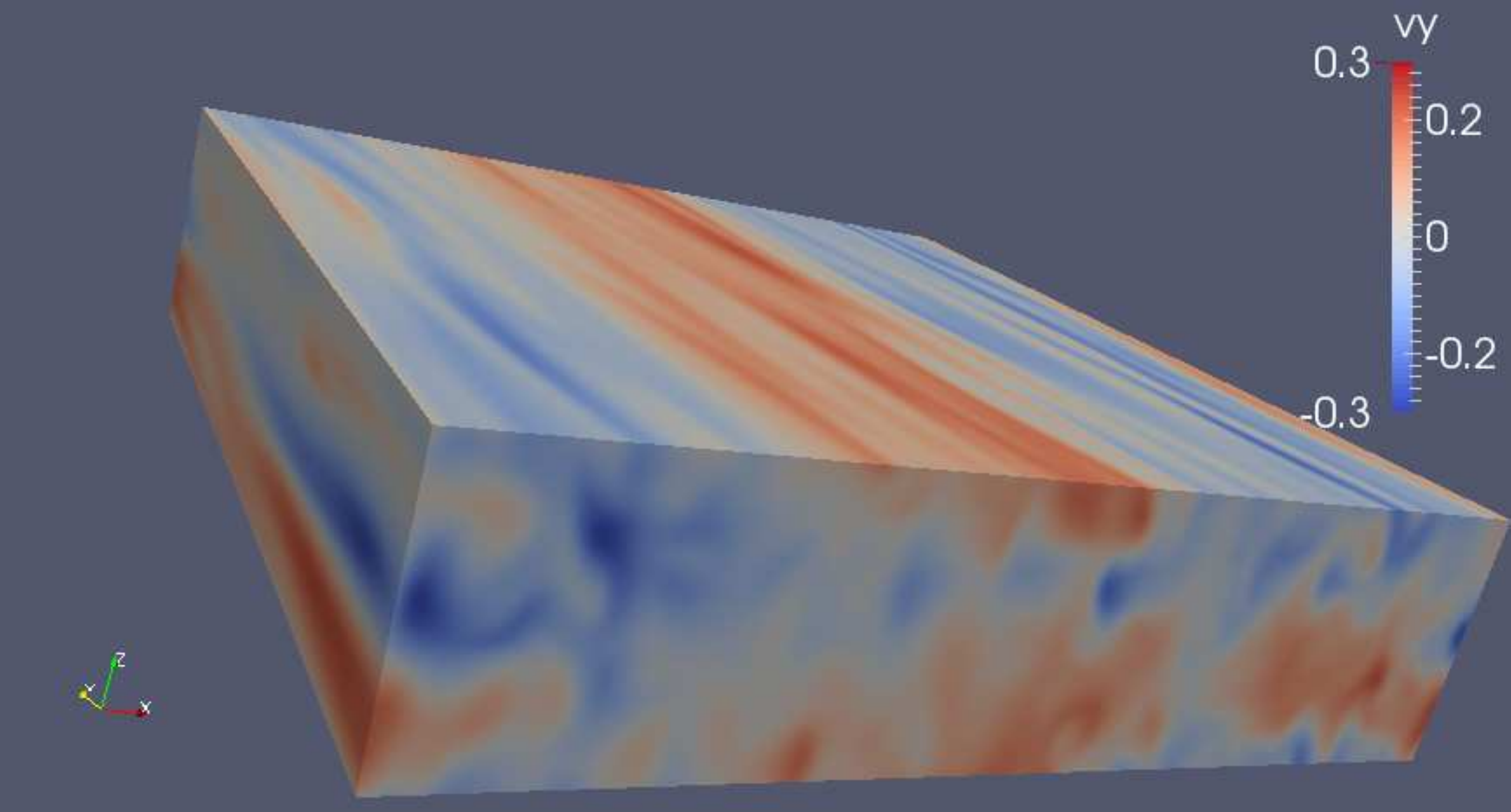} 
 \caption{Snapshots of two simulations of a local model of the MRI in the viscous regime using the Boussinesq approximation (from Guilet \& M\"uller 2015). The left column corresponds to a buoyantly unstable flow with $N^2/\Omega^2 = -1$ while the right column corresponds to a buoyantly stable flow with $N^2/\Omega^2 = 10$. The two rows show the azimuthal magnetic field (top) and azimuthal velocity (bottom). 
 Note that we adjusted the color scales to the level of turbulence, i.e. they are different for the two columns. }
   \label{fig:images_buoyancy}
\end{center}
\end{figure}

\begin{figure}[htbp]
\begin{center}
 \includegraphics[width=6cm]{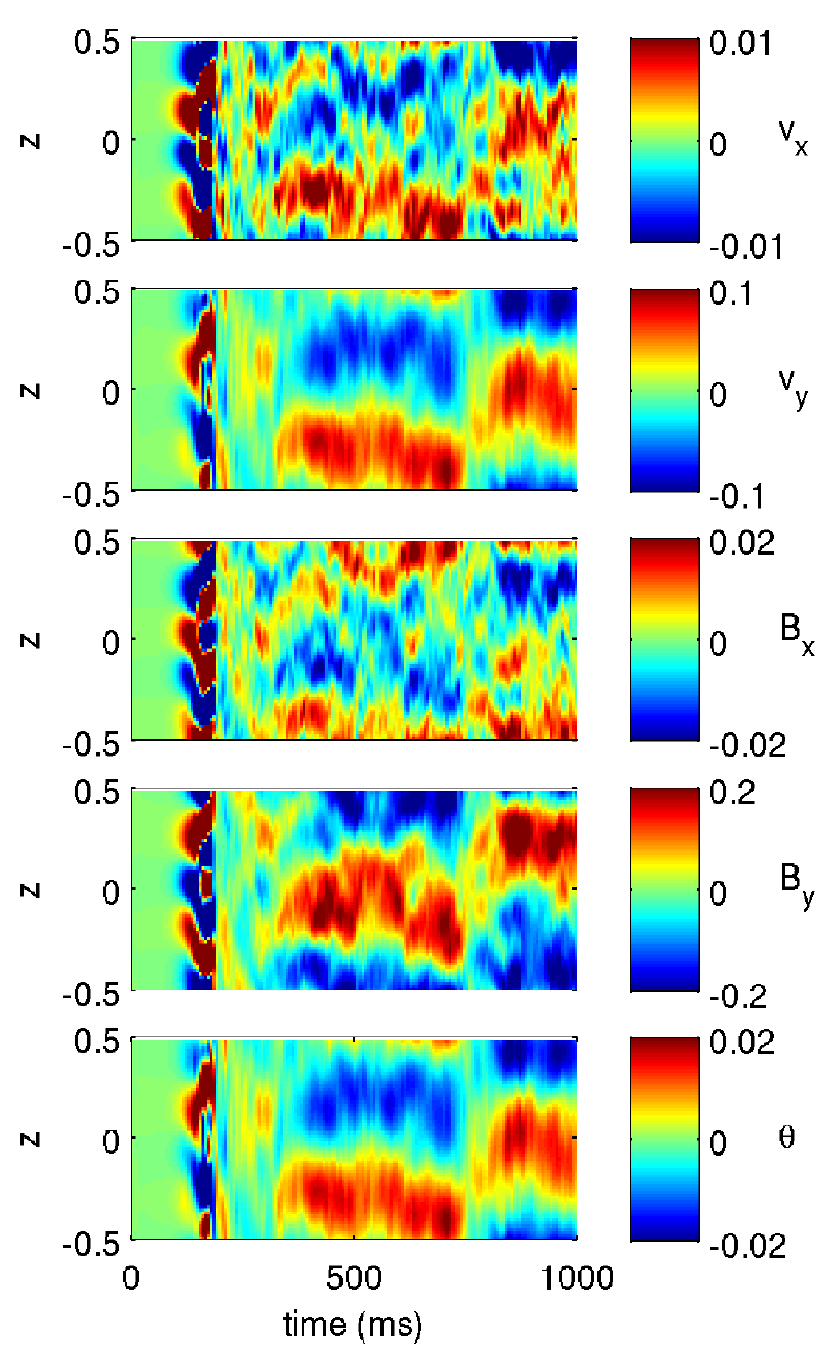} 
 \includegraphics[width=7.4cm]{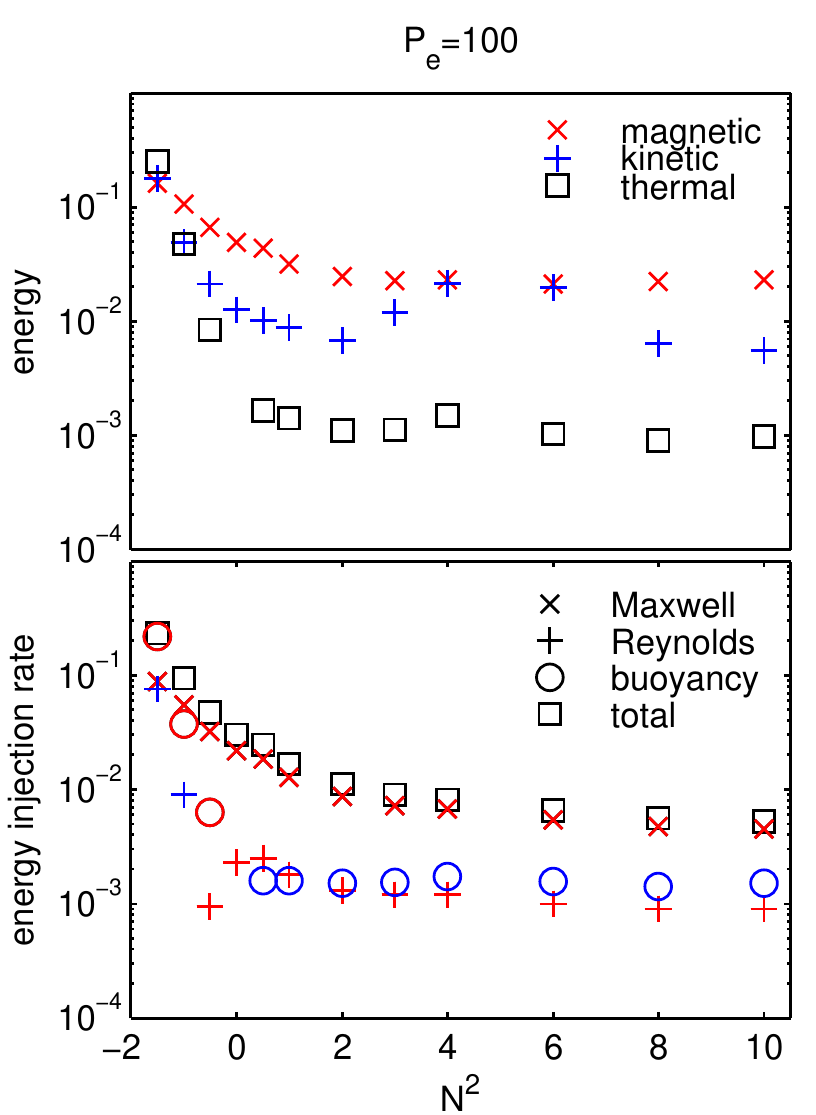} 
 \caption{\textit{Left panel:} Quasi-stationary channel flow for $N^2/\Omega^2=10$ and high thermal diffusion Pe = 100. The five panels show spaceÐtime diagrams (from top to bottom) of the horizontally averaged radial velocity, azimuthal velocity, radial magnetic field, azimuthal magnetic field, and buoyancy variable. \textit{Right panel:} Time and volume averages of the turbulent energy densities (in units of $2\times 10^{29}\,{\rm erg\,cm^{-3}}$), and energy density injection rates (in units of $2\times10^{32}\,{\rm erg\,s^{-1}\,cm^{-3}}$) as a function of the dimensionless buoyancy parameter $N^2/\Omega^2$. They show the magnitude of magnetic (red $\times$), kinetic (blue $+$), and thermal (black squares) energies in the top  panel, and energy injection rates due to Maxwell stress ($\times$
    symbols), Reynolds stress ($+$ signs), buoyancy force work
    (circles), and sum of the three (squares) in the bottom
    panel. Positive injection rates are shown in red color, while
    negative ones (i.e. energy is removed from turbulent motions) are
    shown in blue.  }
   \label{fig:buoyancy_bis}
\end{center}
\end{figure}

In Guilet \& M\"uller (2015), we have studied the turbulent phase following the disruption of the channel modes, taking for the first time into account both the viscosity and diffusion due to neutrinos and the buoyancy force due to entropy/composition gradients. This was made possible owing to the use of the Boussinesq approximation, which we showed to be well suited to a local model of a PNS. In addition to drastically reducing the computing time this approximation avoids artifacts at the boundaries caused by global gradients in fully compressible simulations. Fig.~\ref{fig:images_buoyancy} shows snapshots of the turbulent phase in a buoyantly unstable case (left panel) and a buoyantly stable case (right panel). While the buoyantly unstable dynamics is turbulent and largely non-axisymmetric, the buoyantly stable dynamics contains more large-scale axisymmetric structures. A channel flow at the largest scale available in the box can be distinguished even during the turbulent phase, which is further illustrated in Fig.~\ref{fig:buoyancy_bis} by the horizontally averaged space-time diagram (left panel). Fig.~\ref{fig:buoyancy_bis} (right panel) furthermore shows the kinetic, magnetic and thermal energies and energy injection rates as a function of the dimensionless buoyancy parameter $N^2/\Omega^2$ (where $N$ is the Brunt-V\"ais\"al\"a frequency). The magnetic energy decreases with $N^2/\Omega$ but, interestingly, it becomes almost constant in the stable buoyancy regime ($N^2>0$) suggesting an efficient magnetic field amplification even in the presence of a strong, stable stratification.

While these simulations were able to use realistic values for the viscosity due to neutrinos, they have the shortcoming of vastly overestimating the resistivity (like all other simulations, be it by including it explicitly like here or by suffering from a large numerical resistivity (for estimates see Rembiasz et al. 2016c)). 
Preliminary results show that this is likely to cause a significant underestimate of the efficiency of magnetic field amplification, thus confirming the strong dependence of the MRI on the magnetic Prandtl number (the ratio of viscosity to resistivity) known in the context of accretion disks (e.g. Fromang et al. 2007).

\section{Perspectives}
The presence of structures on the largest scales described by the local models of Section~3 stresses the necessity of global models encompassing the whole PNS. Such models are extremely challenging computationally because it is necessary to resolve the very small length-scales where MRI grows. The closest to a global model was published by M\"osta et al. (2015) who simulated one eighth of a PNS (in an octant symmetry) and claimed to obtain an MRI-driven dynamo. Even this extremely computationally intensive simulation could not, however, answer the fundamental question of the origin of magnetars dipolar magnetic fields because it was actually initialized with a dipolar magnetic field strong enough to lead to magnetar formation by simple flux conservation. More numerical simulations of global models of PNSs are therefore necessary to establish firmly whether the MRI can generate a sufficiently strong dipolar magnetic to explain the birth of millisecond magnetars. We are currently making steps in this direction by developing numerical models of idealized PNSs in quasi-incompressible approximations (Boussinesq or anelastic approximations). By drastically reducing the computational requirements, these approximations should enable insights into the fundamental physical process generating the dipolar magnetic field.   

\acknowledgment{J.G. acknowledges support from the Max-Planck-Princeton Center for Plasma Physics and from the European Research Council (grant MagBURST--715368). H.T.J. is grateful for funding by the European Research Council through grant ERC-AdG COCO2CASA-3411 and by the Deutsche Forschung Gemeinschaft through Cluster of Excellence "Universe" EXC-153. M.A., P.C.D., M.O. and T.R. acknowledge support from the European Research Council (grant CAMAP-259276) and from grants AYA2013-9340979-P, AYA2015-66899-C2-1-P and PROME-TEOII/2014-069. }

\end{document}